\documentclass[11pt
]{amsart}

\usepackage{hyperref}
\hypersetup{bookmarksdepth=3}

\usepackage{geometry}
\usepackage{physics}
\usepackage{amsmath,amssymb}
\usepackage{mathtools}
\usepackage{
mathrsfs}
\usepackage{esint}
\usepackage{braket}
\usepackage{tikz}
\usepackage{pgf}

\usepackage{fancyhdr,lastpage}

\usepackage{graphicx,bbm,multicol,mathtools}
\usepackage{subcaption}
\usepackage{pgfplots}
\usepackage{verbatim}
\usepackage{comment}
\usepackage[normalem]{ulem}

\usepackage{amssymb,amsmath,amsthm}

\theoremstyle{remark}

\usepackage{graphicx}

\newcommand{\R}{{\mathbb R}}

\newcommand{\N}{\mathbbm{N}}

\title[ANALYTICAL SOL.OF A GR-POBLEM WITH TIME-DEPENDENT BOUNDARY CONDITIONS]{ ANALYTICAL SOLUTION OF A GAS RELEASE PROBLEM CONSIDERING PERMEATION WITH TIME-DEPENDENT BOUNDARY CONDITIONS}
\author[Marvin R. Schulz et al.]{ Marvin R. Schulz, Kaori Nagatou, Axel von der Weth, Frederik Arbeiter and Volker Pasler}
\begin{document}
	\begin{abstract}
		In this paper the determination of material properties such as Sieverts' constant (solubility) and diffusivity (transport rate) via so-called gas release experiments is discussed. In order to simulate the time-dependent hydrogen fluxes and concentration profiles efficiently, we make use of an analytical method, namely we provide an analytical solution for the corresponding diffusion equations on a cylindrical specimen and a cylindrical container for three boundary conditions. These conditions occur in three phases -- loading phase, evacuation phase and gas release phase. In the loading phase the specimen is charged with hydrogen assuring a constant partial pressure of hydrogen. Then the gas will be quickly removed by a vacuum pump in the second phase, and finally in the third time interval, the hydrogen is released from the specimen to the gaseous phase, where the pressure increase will be measured by an equipment which is attached to the cylindrical container. The investigated diffusion equation in each phase is a simple homogeneous equation, but due to the complex time-dependent boundary conditions which include the Sieverts' constant and the pressure, we transform the homogeneous equations to the non-homogeneous ones with a zero Dirichlet boundary condition. Compared with the time consuming numerical methods our analytical approach has an advantage that the flux of desorbed hydrogen can be explicitly given and therefore can be evaluated efficiently. Our analytical solution also assures that the time-dependent boundary conditions are exactly satisfied and furthermore that the interaction between specimen and container is correctly taken into account.

		

	\end{abstract}
\maketitle \noindent
This is the original manuscript of the main author. 
This article has been accepted for publication in the Journal of Computational and Theoretical Transport , published by Taylor \& Francis. The accepted manuscript is available on the web page of the main author \href{https://www.math.kit.edu/iana1/~schulz/de}{https://www.math.kit.edu/iana1/~schulz/de}.
\newpage
\tableofcontents



\section{INTRODUCTION}
The diffusion of hydrogen in metals plays a role in metallurgy, vacuum technology and many disciplines of process and energy engineering. In the field of nuclear fusion (where this present work originates), hydrogen isotopes are the fuel for the energy producing fusion reaction. Consequently, hydrogen is present in many parts (inside and outside of the fusion core) of present day fusion experiments and future power plants. The interaction of hydrogen with the contacted equipment in terms of absorption and permeation relates to safety, fuel budget and lifetime of the components. 
The present work is part of an endeavor to complete the available foundation of data and modelling theories on the diffusion and permeation of hydrogen isotopes in steel at fusion-relevant conditions. Characteristic conditions are: low hydrogen partial pressures ($1-1000 \, \text{Pa}$), co-permeation of hydrogen isotopes (H, D, T), manufacturing and heat treatment techniques which modify surface and grain structure, and finally the effects of energetic neutron irradiation on the material lattice and transmutation.
Experiments are needed to determine material properties such as the Sieverts' constant ($k_s$) and Diffusivity ($D$) as function of relevant conditions (temperature, partial pressure) for various material processing histories, including neutron irradiation. So-called gas release experiments are studied as promising technique to investigate relatively small irradiated steel specimens.
The objective of the presented work is to provide efficient methods to simulate the time-dependent hydrogen fluxes $j(t)$ and concentration fields $c(r,z,t)$ of such experiments. While numerical methods such as finite differences (FDM) \cite{VDWA2019} or finite volume methods (FVM) \cite{VP2018} allow the analysis of arbitrary 3D geometries, their numerical cost is relatively high (hours to days per single run). On the other hand, there is demand for fast and efficient execution of large number of runs: First, for parameter-variation studies during the experimental design phase to optimize setup and boundary conditions for uncertainty reduction, and second for derivation of the transport parameters \{$D$, $k_s$\} from measured experimental signals by iterative methods such as the branch-and-bound algorithm \cite{VDWA201902}. In both cases, hundreds or thousands of runs are needed per task.
Therefore, we developed analytical solutions to describe the time-dependent concentration profiles and surface fluxes for specimens and components of a gas-release experiment in 2D cylindrical coordinates $(r,z)$.
\begin{figure}[h]
	\begin{subfigure}[t]{0.6\textwidth}
		\includegraphics[width=0.9\textwidth]{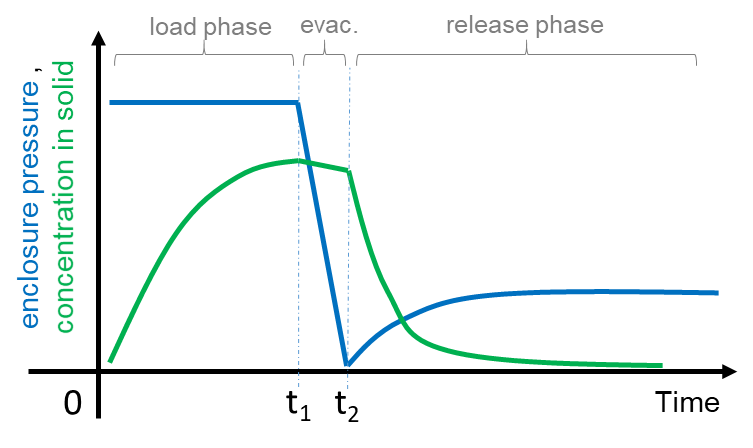}
		\caption{Pressure curve during the experiment.} \label{prs_curve}
	\end{subfigure}
	\hfill
	\begin{subfigure}[t]{0.37\textwidth}
		\includegraphics[width=0.9\textwidth]{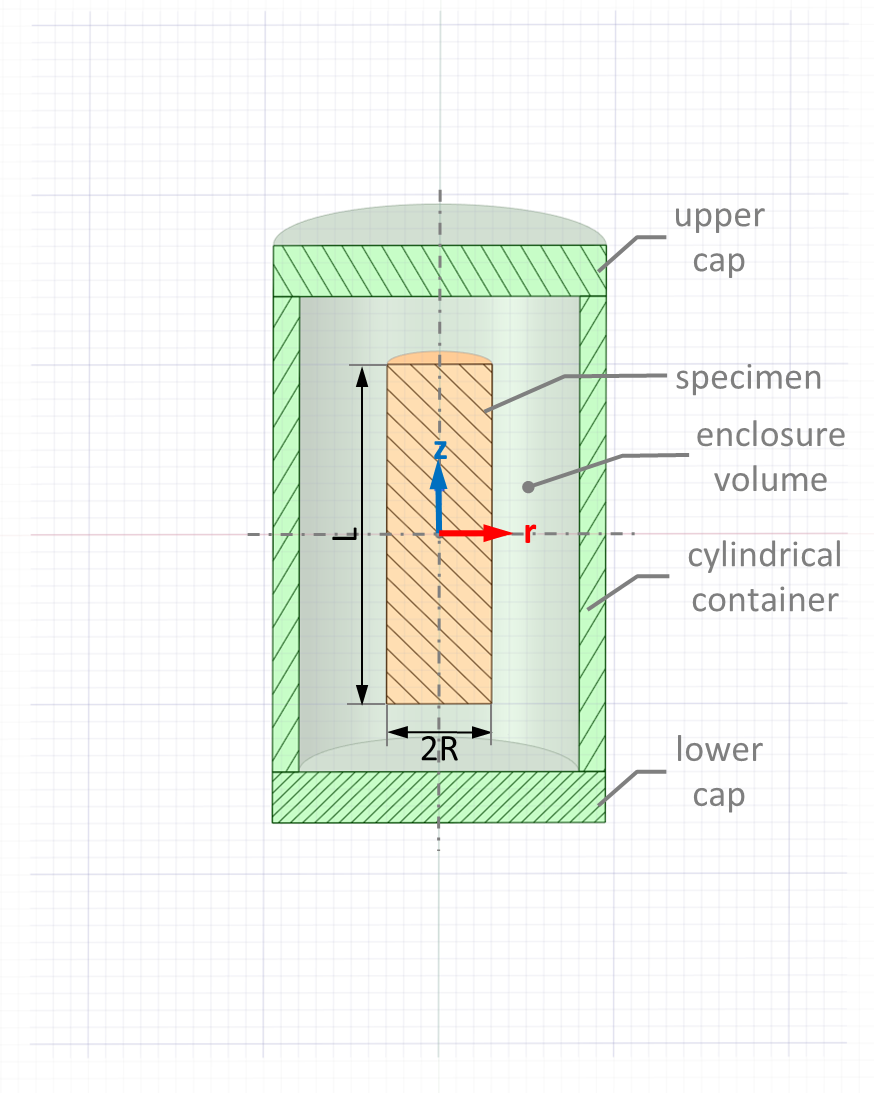}
		\caption{Setup of the experiment.}  \label{prs_setup}
	\end{subfigure}
	\caption{Gas Release Experiment.\label{Fig1}}
\end{figure}
\section{Gas Release Experiments} \label{gas_rel_int}
In the considered gas release experiments, cylindrical specimens/samples are placed in a hermetic enclosure formed by an annular can with upper and lower disk-like caps (See  Figure \ref{prs_setup}). The enclosure is attached to process equipment, namely a pressure measurement device, a hydrogen gas supply and a vacuum pump. We assume isothermal conditions for the following mathematical derivations (non-isothermal corrections can be easily applied). A gas release experiment is run in the following phases (see Figure \ref{prs_curve}):
\begin{enumerate}
	\item \textbf{Loading phase ($t_0=0 \leq t \le t_1$): }the gas supply system assures a constant partial pressure of hydrogen $p_L$ in the enclosure throughout the loading time (This is realized by constantly flushing the enclosure with a mixture of inert gas and hydrogen at constant total pressure). During this phase, hydrogen is solved in the specimen (and the enclosure walls) and concentration profiles evolve.
	\item \textbf{Evacuation ($t_1 \leq t \le t_2$): }the gas is pumped out of the enclosure (very fast) down to a residual level $p_0$. The pumping interval must be limited, because some hydrogen is already lost from the specimen during the pumping.
	\item \textbf{Gas release phase ($t \geq t_2$): }hydrogen is released from the specimen (and the enclosure walls) to the gas phase, until an equilibrium between the $H_2$ in the gas phase and the concentration of the solute hydrogen atoms $H$ in the specimen’s metal lattice satisfies the Sieverts' law, $c_H = k_s  \sqrt{p_{H_2}}$. The hydrogen flux from the surface $\dot{J}_{H}(t)$ accumulates in the enclosure volume as molecular hydrogen $ n_{H_2}(t)= \frac{1}{2} \int_{\tau=0}^{t} \dot{J}_{H}(\tau)d\tau $, and builds up the pressure $p_{H_2}(t)=\frac{n(t)\cdot \textit{\textbf{R}}T}{V_{e}}$, which is the primary measurement signal of a gas release experiment.\footnote{{In the following sections, all pressures refer to molecular gas, i.e. $H_2$, and all concentrations refer to solute atomic hydrogen. The indices $H_2$ and $H$ are usually dropped for better readability.}}
\end{enumerate}
In a simplified view, the rise-time of a gas release experiment contains the information on the diffusivity, while the steady state pressure level relates to the Sieverts’ constant. However, for realistic experimental setups, it is necessary to consider the hydrogen flux budget of all involved components and volumes. The presented approach therefore includes the wall of the can, which is an inevitable contributor to the hydrogen budget.
\subsection{Analytic Model of the Gas Release Experiment}

The Diffusion Equation is a well understood partial differential equation. Still, the exact solution to special geometries may be hard to find or even does not exist at all. In order to describe the given experiment one struggles with time-dependent boundary conditions coupling the Diffusion Equation in several bodies. In this section we provide the exact solution in the first two time intervals to the problem. We suggest an approach to the exact solution in the third time interval.\\
We extend the existing model by Sedano et al. in \cite{SL1999} for the two dimensional case and in addition we apply the principle to the surrounding container. Note that there is already an extensive discussion of the experiment and on a numerical approach using some Finite Difference Method in \cite{VDWA2019}. See \cite{KP2016} for a detailed discussion on a problem considering heat conduction in a cylindrical solid. 
\subsection{Diffusion Equation with Time-Dependent Boundary Conditions} \label{equations}
We assume an idealized cylindrical geometry for the experimental setup and discuss the explicit solution to the Diffusion Equation also known as Heat Equation for different time-dependent boundary conditions on a cylinder. \\
The concentration distribution in the specimen in the time interval $[t_i,t_{i+1}]\subset \R_+$ is given by 
\begin{align} \label{heat_specimen}
\begin{cases}
\partial_t c_i \left( \Phi (\textbf{r}),t \right) - D_s \Delta c_i\left( \Phi (\textbf{r}),t \right)  = 0 \quad &\textbf{r} \in \mathcal{U}, t \in [t_i,t_{i+1}],	\\
c_i\left( \Phi (\textbf{r}),t \right) = k_s \sqrt{p_i(t)} &\textbf{r} \in \partial \mathcal{U},t \in [t_i,t_{i+1}],\\ 
c_i\left( \Phi (\textbf{r}),t_i \right)= c_{i-1}\left( \Phi (\textbf{r}),t_i \right)&\textbf{r} \in  \mathcal{U}.
\end{cases}
\end{align}
Here $\Phi$ is the transformation for cylindrical coordinates. The set $\mathcal{U}= \Phi^{-1}\left(U \times [0,2\pi) \right) \subset \R^3$ is a compact set describing the specimen where  $U =[0,R]\times[\frac{-L}{2},\frac{L}{2}] $. We assume $t_0=0$ and $t_i<t_{i+1}$ for $i \in \{0,1,2\}$ and $p_i: [t_i,t_{i+1}] \to \R$ to be the pressure of the gaseous phase. The constants $D_s$ and $k_s$ are positive real numbers. We assume that $c_0\left( \Phi (\textbf{r}),t_0 \right)\equiv 0$. \\
The concentration distribution in the container can be described with
\begin{align} \label{heat_container}
\begin{cases}
\partial_t u_i\left( \Phi (\textbf{r}),t \right) - D_c \Delta u_i\left( \Phi (\textbf{r}),t \right) = 0& \textbf{r} \in \mathcal{V}, t \in [t_i,t_{i+1}],	\\
u_i\left( \Phi (\textbf{r}),t \right) = k_s^{(c)} \sqrt{p_i(t)} &\textbf{r} \in \partial\mathcal{V}_{-},t \in [t_i,t_{i+1}],\\ 
u_i\left( \Phi (\textbf{r}),t \right) = 0 &\textbf{r} \in \partial \mathcal{V}_{+},t \in [t_i,t_{i+1}],\\ 
u_i\left( \Phi (\textbf{r}),t_i \right)= u_{i-1}\left( \Phi (\textbf{r}),t_i \right) &\textbf{r} \in  \mathcal{V}.
\end{cases}
\end{align}
Here $\mathcal{V} = \Phi^{-1}(V\times \R \times [0,2\pi) ) \subset \R^3$ with $V=[R_1,R_2]$ is a set describing the container, $\partial \mathcal{V}_{-}$ denotes the inner surface and $\partial \mathcal{V}_{+}$ the outer surface. The constants $D_c$ and $k_s^{(c)}$ are positive real numbers. For simplification we shall assume that the container is an infinitely long hollow cylinder.\\
Firstly it is convenient to change to cylindrical coordinates. Note that the Laplace Operator reads
\begin{align} \label{lap_in_cyl}
\Delta = \frac{1}{r} \partial_r + \partial_r^2 + \frac{1}{r^2} \partial_\varphi^2 + \partial_z^2. 
\end{align}
By some symmetrie arguments we have
\begin{align}
&\partial_z u_i \equiv \partial_{\varphi} u_i \equiv 0, \,\, \partial_\varphi c_i \equiv 0,   \\
&\partial_z c_i(r,0,\varphi, t)=0 \quad \forall  \, (r,\varphi) \in [0,R] \times [0,2\pi) \text{ and } t\in \R_+, \\
&\partial_r c_i(0,z,\varphi, t)=0 \quad \forall  \, (z,\varphi) \in \left[\frac{-L}{2},\frac{L}{2}\right] \times [0,2\pi) \text{ and } t\in \R_+, \\
& \partial_z u_i(r,0,\varphi,t) = 0  \quad \forall  \, (z,\varphi) \in \R \times [0,2\pi) \text{ and } t\in \R_+, 
\end{align}
and hence it is convenient to assume $c_i: U \to \R$ and $u_i:V \to \R$. For sufficiently smooth $p$ the solution to \eqref{heat_specimen} and \eqref{heat_container} exists as we conclude below and is given using Duhamel's principle. See for explanation \cite{JJ1998}[Lemma 4.3.4].  \\
In order to apply Duhamel's formula it is necessary to transform \eqref{heat_specimen} and \eqref{heat_container} for zero boundary conditions. Thus, we define
\begin{align} \label{transform_1}
&g_i(r,z,t) \coloneqq c_i(r,z,t) - k_s \sqrt{p_i(t)},\\
&h_i(r,t) \coloneqq u_i(r,t) - \omega(r) k_s^{(c)} \sqrt{p_i(t)}, \label{cast_u_h}
\end{align}
for all $(r,z) \in U$ respectively $r \in V$ and $t \in [t_i,t_{i+1}]$. Since the boundary condition at the edge of the container for $r=R_1$ differs from the condition at $r=R_2$ one has to interpolate between both sides with some function $\omega: [R_1,R_2] \to [0,1]$ with $\omega(R_1) = 1$ and $\omega(R_2) =0$. As in \cite{KP2016} suggested, it is convenient to use the function
\begin{align}
\omega(r) \coloneqq \frac{\log(r) - \log(R_2)}{\log(R_1) - \log(R_2)}
\end{align}
since the logarithm is the Green's function to the radial Laplace operator and thus $\Delta \omega \equiv 0$. The functions $g_i,h_i$ fulfill the corresponding non-homogeneous equations
\begin{align} \label{heat_specimen_reshape}
\begin{cases}
\partial_t g_i(r,z,t) - D_s \Delta g_i(r,z,t) = -k_s \partial_t \sqrt{p_i(t)} \eqqcolon f_{i,s}(t) \quad &(r,z) \in U,  t \in [t_i,t_{i+1}],	\\
g_i(r,z,t) =0 & (r,z) \in \partial U,t \in [t_i,t_{i+1}],\\ 
g_i(r,z,t_i)= g_{i-1}(r,z,t_{i}) &(r,z) \in U,
\end{cases}
\end{align}
and
\begin{align} \label{heat_container_reshape}
\begin{cases}
\partial_t h_i(r,t) - D_c \Delta h_i(r,t) = \omega(r) f_{i,v}(t) \quad &r \in V,  t \in [t_i,t_{i+1}],	\\
h_i(r,t) =0 & r \in \partial V,t \in [t_i,t_{i+1}],\\ 
h_i(r,t_i)= h_{i-1}(r,t_{i}) &r \in V.
\end{cases}
\end{align}
The solutions to these non-homegeneous Heat Equations are well known for a large class of functions $f_{i,v}(t), f_{i,s}(t)$. See for example \cite{SA1947}[\S\,20] for the isobaric case $\partial_t f_{i,v} = \partial_t f_{i,s} = 0$.
\subsection{Duhamel's Formula}
The solution to the Equations \eqref{heat_specimen_reshape} and \eqref{heat_container_reshape} is given by Duhamel's formula
\begin{align} \label{duhamel_1}
g_i(r,z,t+t_{i-1}) &= \exp[D_s\Delta t]g_{i-1}(r,t_{i-1})+\int_{0}^t \exp[D_s\Delta (t-s)] f_{i,s}(s) ds, \\
h_i(r,t+t_{i-1}) &= \exp[D_c\Delta t]h_{i-1}(r,t_{i-1})+\int_{0}^t \exp[D_c\Delta (t-s)]  \omega(r) f_{i,v}(s) ds. \label{duhamel_2}
\end{align}
Here we use the operator valued representation of the heat kernel $\exp[D\Delta t]$, that can be extended using the spectral decomposition of $\Delta$ on $U$ respectively $V$. Denote the eigenfunctions of $\Delta$ on $U$ with zero boundary conditions as $\varphi_{m,n}$ and the eigenvalues as $-\lambda_{m,n}^2$. Using standard techniques one finds
\begin{align} \label{eigenfunc_s}
&\varphi_{m,n}(r,z) = \frac{2}{R \mathcal{J}_1(x_m)} \mathcal{J}_0\left( x_m \frac{r}{R} \right) \cos\left(\frac{(2n+1)\pi}{L}z\right) &(m,n)\in \N_0^2, \\
&\lambda_{m,n}^2 = \left(\frac{x_m}{R}\right)^2 + \left(\frac{(2n+1)\pi}{L}\right)^2 &(m,n)\in \N_0^2.  \label{Eigenval_s}
\end{align}
Here $\mathcal{J}_0$ is the zero order Bessel function of the first kind and $x_m$ its $m$-th zero point. Denote the eigenfunctions of $\Delta$ on $V$ with zero boundary conditions as $\psi_{n}$. Using standard techniques one finds
\begin{align}\label{eig_container}
\psi_{n}(r) = \frac{-Y_0(y_n R_1)}{\mathcal{J}_0(y_n R_1)} \mathcal{J}_0(y_n r) + Y_0(y_n r), \quad n \in \N_0
\end{align}
with $Y_0$ the zero order Bessel function of the second kind, sometimes referred as Neumann's function, and $y_n$ the $n$-th zero point of the polynomial,
\begin{align}
\det \begin{pmatrix} \label{polynom}
\mathcal{J}_0(y R_1 ) & Y_0(y R_1 )  \\
\mathcal{J}_0(y R_2) & Y_0(y  R_2) 
\end{pmatrix} =\mathcal{J}_0(y R_1 )Y_0(y  R_2)-\mathcal{J}_0(y R_2)Y_0(y R_1 ).
\end{align}
Note that the functions $\psi_{n}$ are not normalized until now. The corresponding eigenvalues are $\omega_n = -y_n^2$. The zero point $y_n$ can either computed numerically or estimated using the asymptotic behavior of Bessel's and Neumann's functions as
\begin{align}
y_n \approx \frac{n \pi}{R_2-R_1}, \quad n \in \N.
\end{align} 
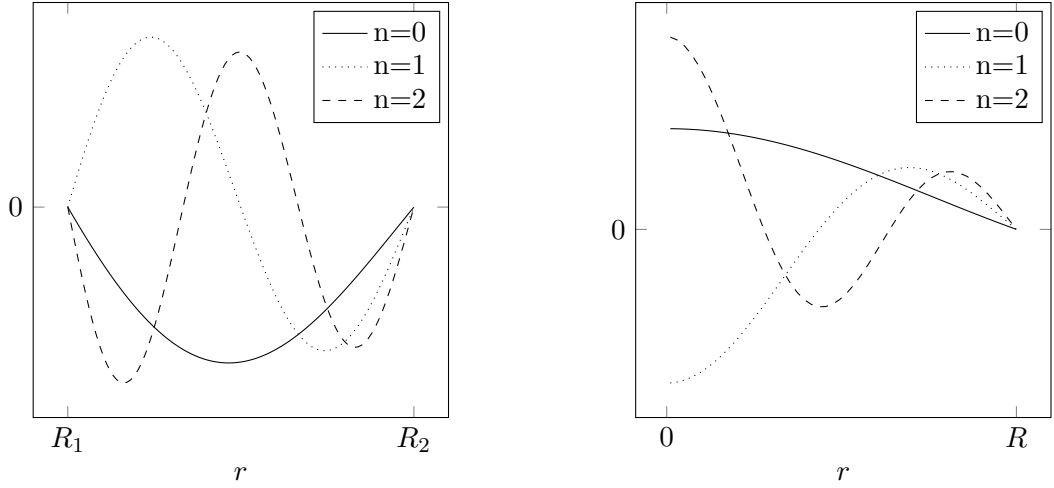
\begin{figure}[h]
	\begin{subfigure}[t]{0.47\textwidth}
		\begin{tikzpicture}
		\begin{axis}[
		scaled ticks=false,
		width=\textwidth, 
		height=\textwidth, 
		xtick={0.01, 0.02},
		xlabel={$r$},
		xticklabels={$R_1$, $R_2$},
		ytick={0}
		]
		\addplot[color=black] table [x=a, y=b, col sep=comma] {data0};
		\addplot[color=black, dotted] table [x=a, y=b, col sep=comma] {data1};
		\addplot[color=black, dashed] table[x=a, y=b, col sep=comma] {data2};
		\addlegendentry{n=0}
		\addlegendentry{n=1}
		\addlegendentry{n=2}
		\end{axis}
		\end{tikzpicture}
		\caption{Normalized eigenfunctions $\psi_{n}$, defined in Equation \eqref{eig_container}. Here we chose $R_1=0.01 \, \text{m}$ and $R_2=0.02 \, \text{m}$.} \label{pic_vess_eig}
	\end{subfigure}
	\hfill
	\begin{subfigure}[t]{0.47\textwidth}
		\begin{tikzpicture}
		\begin{axis}[
		scaled ticks=false,
		width=\textwidth, 
		height=\textwidth, 
		xlabel={$r$},
		xtick={0.0 ,0.003},
		xticklabels={$0$ , $R$},
		ytick={0}
		]
		\addplot[color=black] table [x=a, y=b, col sep=comma] {data0_s};
		\addplot[color=black, dotted] table [x=a, y=b, col sep=comma] {data1_s};
		\addplot[color=black, dashed] table[x=a, y=b, col sep=comma] {data2_s};
		\addlegendentry{n=0}
		\addlegendentry{n=1}
		\addlegendentry{n=2}
		\end{axis}
		\end{tikzpicture}
		\caption{Normalized eigenfunctions $\varphi_{0,n}$ for $z\equiv0$, defined in Equation \eqref{eigenfunc_s}. Here, $R=0.003 \, \text{m}$.}  \label{pic_spec_eig}
	\end{subfigure}
	\caption{Some eigenfunctions in both domains.}
\end{figure}
\subsection{Solution for each Time Interval}
Using the Equation \eqref{duhamel_1} and the eigenfunctions and values assigned in the Equations \eqref{eigenfunc_s} and \eqref{Eigenval_s} the solution to Equation \eqref{heat_specimen_reshape} in each time interval is given by
\begin{align}\label{sol_1}
g_i = \sum_{m,n} &\Big( \exp[-\lambda_{m,n}^2 D_s t] \langle  \varphi_{m,n}, g_{i-1}|_{t=t_{i-1}}  \rangle  \\
& + \int \exp[-\lambda_{m,n}^2 D_s (t-s)] \langle\varphi_{m,n}, f_{i,s}(s)  \rangle ds \Big)   \varphi_{m,n}. \nonumber
\end{align}
Here the product $\langle \cdot,\cdot \rangle$ is given by the orthogonality relation of Bessel's functions as
\begin{align}
\langle f,g \rangle =  \iint_{[0,R]\times[-\frac{L}{2},\frac{L}{2}]} f(r,z) g(r,z) r \frac{d(r,z)}{ L}.
\end{align}
Thus one has $\langle \varphi_{m,n},\varphi_{k,l} \rangle = \delta_{km} \delta_{ln}$. Note that the function \eqref{eigenfunc_s} is indeed normalized, as we show in \ref{coefficients}. The spectral decomposition of $\Delta$ is an orthogonal family in $L^2$. Hence, the Equation \eqref{sol_1} will converge in a $L^2$ sense if the functions $g_{i-1},f_{i,s}$ are regular. Analogously one can derive for the container
\begin{align}\label{sol_2}
h_i = \sum_{n} &\Big( \exp[-y_n^2 D_c t] \frac{\langle  \psi_n, h_{i-1}|_{t=t_{i-1}}  \rangle}{\abs{\langle  \psi_n, \psi_n  \rangle}} + \int \exp[-y_n^2 D_c (t-s)]  \frac{\langle\psi_n, \omega f_{i,v}(s)  \rangle}{\abs{\langle  \psi_n, \psi_n  \rangle}} ds \Big)   \psi_{n}.
\end{align}
Note that in this one dimensional case the inner product is,
\begin{align} \label{inner_prod_2}
\langle f,g \rangle =  \int_{[R_1,R_2]} f(r) g(r) r dr.
\end{align}
In the first time interval one has $p_1(t) \equiv p_L$ the loading pressure and $u_0(r,0) = c_0(r,0) \equiv 0$. Thus, one can compute all of the remaining integrals. Using the derived solutions $u_1,c_1$ and the pressure in the second interval $p_2(t) = \exp[-\sigma t] p_L$ for some $\sigma>0$ one can find the full solution in the second interval.\\
In order to use the Equations $\eqref{sol_1}$ and $\eqref{sol_2}$ in the last time interval one has to propose an ansatz for $p_3(t)$. We use the ansatz proposed by Sedano et. al in \cite{SL1999}
\begin{align} \label{sed_an}
\sqrt{p_3(t+t_2)} =  \frac{\sqrt{p_f} - \sqrt{p_0}}{1-\exp[- \beta \tau]}\left( 1-\exp[-\beta t] \right) + \sqrt{p_0}.
\end{align}
Here $\beta,p_f$ and $\tau$ are free parameters used to optimize the solution and $p_0$ is the residual pressure (as explained in Section \ref{gas_rel_int}). We solve all of the remaining spatial integrals in \ref{coefficients}[Eq. \eqref{t_1} and Eq. \eqref{t_2}] and give the time integrals in the following. In conclusion one has 
\begin{align} \label{t_1_g_sol}
g_1(r,z,t) &= \sum_{m,n}(-1)^{n+1}\frac{8k_s \sqrt{p_L}\exp[-\lambda_{m,n}^2 D_s t]}{x_m \mathcal{J}_1(x_m)(2n+1) \pi}  \mathcal{J}_0\left( x_m \frac{r}{R} \right) \cos\left(\frac{(2n+1)\pi}{L}z\right), \\\label{t_1_h_sol}
h_1(r,t) &= \sum_n \frac{-k_s^{(c)}\sqrt{p_L} \langle  \psi_n,\omega  \rangle}{\abs{\langle  \psi_n, \psi_n  \rangle}}\exp[-y_n^2 D_c t]\left(\frac{-Y_0(y_n R_1)}{\mathcal{J}_0(y_n R_1)} \mathcal{J}_0(y_n r) + Y_0(y_n r)\right).
\end{align}
\begin{figure}[h]
	\centering
	\begin{subfigure}[b]{0.47\textwidth}
		\begin{tikzpicture}
		\begin{axis}[
		scaled ticks=false,
		width=\textwidth, 
		height=\textwidth, 
		xlabel={$r$},
		xtick={0.0 ,0.003},
		ytick={0.0,1.0},
		xticklabels={$0$ , $R$},
		yticklabels={$0$ , $k_s \sqrt{p_L}$}
		]
		\addplot[color=black] table [x=a, y=b, col sep=comma] {data0_s_Load};\label{p0}
		\addplot[color=black,dashed] table [x=a, y=b, col sep=comma] {data1_s_Load};\label{p1}
		\addplot[color=black,dashed] table[x=a, y=b, col sep=comma] {data2_s_Load};\label{p2}
		\addplot[color=black,dashed] table [x=a, y=b, col sep=comma] {data3_s_Load};\label{p3}
		\addplot[color=black,dashed] table [x=a, y=b, col sep=comma] {data4_s_Load};\label{p4}
		\addplot[color=black,dotted] table[x=a, y=b, col sep=comma] {data5_s_Load};\label{p5}
		\end{axis}
		\end{tikzpicture}
		\caption{Solution $c_1$ for $z= 0$ in the first time interval.} \label{pic_spec_sol1}
	\end{subfigure}
	\hfill
	\begin{subfigure}[b]{0.47\textwidth}
		\begin{tikzpicture}
		\begin{axis}[
		scaled ticks=false,
		width=\textwidth, 
		height=\textwidth, 
		xtick={0.01, 0.015},
		ytick={0.0, 1.0},
		xlabel={$r$},
		xticklabels={$R_1$, $\frac{R_1+R_2}{2}$},
		yticklabels={$0$ , $k_s^{(c)} \sqrt{p_L}$},
		]
		\addplot[color=black] table [x=a, y=b, col sep=comma] {data0_v_Load};
		\addplot[color=black,dashed] table [x=a, y=b, col sep=comma] {data1_v_Load};
		\addplot[color=black,dashed] table[x=a, y=b, col sep=comma] {data2_v_Load};
		\addplot[color=black,dashed] table [x=a, y=b, col sep=comma] {data3_v_Load};
		\addplot[color=black,dotted] table [x=a, y=b, col sep=comma] {data4_v_Load};
		\end{axis}
		\end{tikzpicture}
		\caption{Solution $u_1$ in the first time interval.} \label{pic_ves_sol1}
	\end{subfigure} 
	\caption{Solutions in the first time interval as given by the Equations \eqref{t_1_g_sol} and \eqref{t_1_h_sol}. Each graph belongs to one defined time and (\ref{p0}, $t=0 \, \text{s}$) and (\ref{p5}, $t=800 \, \text{s}$). The graphs (\ref{p2}) belong to times $0\, \text{s}<t<800\, \text{s}$. We used $R=3.0e^{-3} \, \text{m}$, $k_s=1.829e^{-3}\frac{\text{mol}}{\text{m}^3 \sqrt{\text{Pa}}}$, $L= 6.0e^{-2}\, \text{m}$, $p_L=3.0e^3 \, \text{Pa}$, $D_s=7.879e^{-9} \, \frac{\text{m}^2}{\text{s}}$, $R_1=1.0e^{-2} \, \text{m}$, $R_2=2.0e^{-2}  \, \text{m}$, $k_s^{(c)}=5.914e^{-4}\frac{\text{mol}}{\text{m}^3 \sqrt{\text{Pa}}}$ and $D_c= 8.257e^{-10}\, \frac{\text{m}^2}{\text{s}}$.} \label{fig_t_1}
\end{figure}
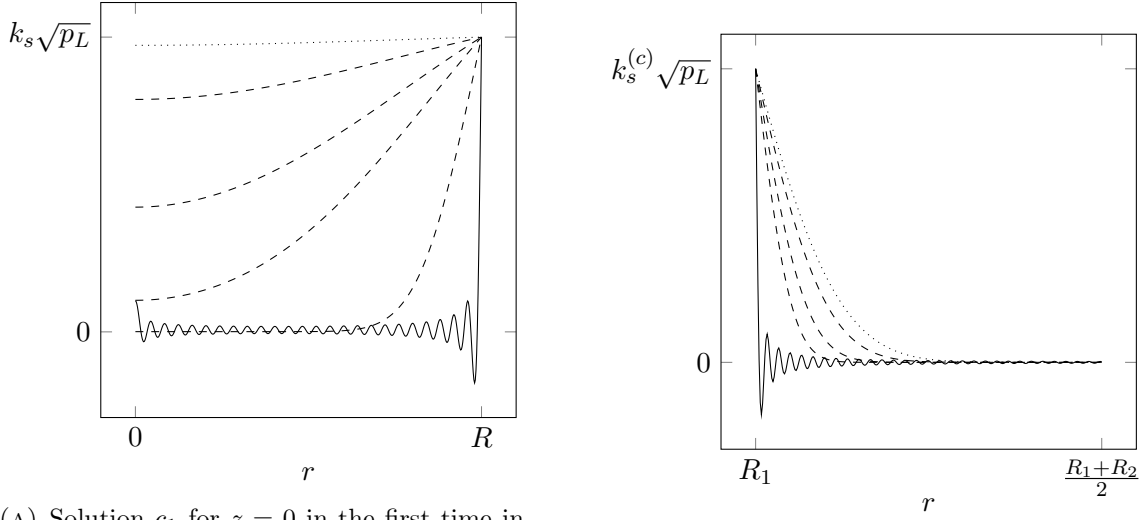 
The corresponding solutions $c_1$ and $u_1$ are plotted in Figure \ref{fig_t_1}. As explained one uses this solution in the first time interval to find the solution in the second interval. The homogeneous part of Duhammel's formula can be simply recovered from the previous case. One concludes directly
\begin{align}
&\langle \varphi_{mn},g_1|_{t=t_1} \rangle =-k_s\sqrt{p_L} \langle \varphi_{mn},1 \rangle \exp\left[-\lambda_{m,n}^2D_st_1\right],\\
&\langle \psi_{n},h_1|_{t=t_1} \rangle = -k_s^{(c)}\sqrt{p_L} \langle  \psi_n,\omega  \rangle \exp\left[-y_n^2D_ct_1\right].
\end{align}
In addition to the previous case one has to find the specific solution. Thus, one computes
\begin{align}
&\int_0^t \exp[-\lambda_{m,n}^2 D_s (t-s)] \langle\varphi_{m,n}, f_{2,s}(s) \rangle ds  \\
= &\frac{\sigma k_s \sqrt{p_L}}{2\lambda_{m,n}^2 D_s - \sigma } \langle \varphi_{mn},1 \rangle \underset{\eqqcolon \nu_{m,n}(t)}{\underbrace{\left( \exp\left[-\frac{\sigma}{2}t\right] -  \exp\left[\lambda_{m,n}^2D_s t\right] \right)}}. \nonumber
\end{align}
For the container the same calculation holds
\begin{align}
&\int_0^t \exp[-y_n^2 D_c (t-s)]  \frac{\langle\psi_n, \omega f_{2,v}(s)  \rangle}{\abs{\langle  \psi_n, \psi_n  \rangle}} ds   \\
=&  \frac{\sigma \langle\psi_n, \omega \rangle k_s^{(c)} \sqrt{p_L}}{\abs{\langle  \psi_n, \psi_n  \rangle}(2y_n^2 D_c - \sigma)} \underset{\eqqcolon \kappa_{n}(t)}{\underbrace{ \left( \exp\left[-\frac{\sigma}{2}t\right] -  \exp\left[y_n^2D_ct\right] \right)}}. \nonumber
\end{align}
Hence, the solutions in the second interval read
\begin{align}
&g_2(r,z,t+t_1) = \sum_{m,n}(-1)^{n+1}\frac{8k_s \sqrt{p_L}}{x_m\mathcal{J}_1(x_m) (2n+1) \pi} \label{t_2_g_sol} \\
&\times \left(\exp[-\lambda_{m,n}^2 D_s (t+t_1)]-\frac{\sigma }{2\lambda_{m,n}^2 D_s - \sigma } 
\nu_{m,n}(t)\right) \nonumber \mathcal{J}_0\left( x_m \frac{r}{R} \right) \cos\left(\frac{(2n+1)\pi}{L}z\right), \\
&h_2(r,z,t+t_1) = \sum_{n}\frac{-k_s^{(c)}\sqrt{p_L} \langle  \psi_n,\omega  \rangle}{\abs{\langle  \psi_n, \psi_n  \rangle}} \left(\exp[-y_n^2 D_s (t+t_1)]-\frac{\sigma  \kappa_{n}(t)}{\left(2y_n^2 D_c - \sigma\right) } \right) \label{t_2_h_sol} \\ \nonumber
&\times \left(\frac{-Y_0(y_n R_1)}{\mathcal{J}_0(y_n R_1)} \mathcal{J}_0(y_n r) + Y_0(y_n r)\right).
\end{align}
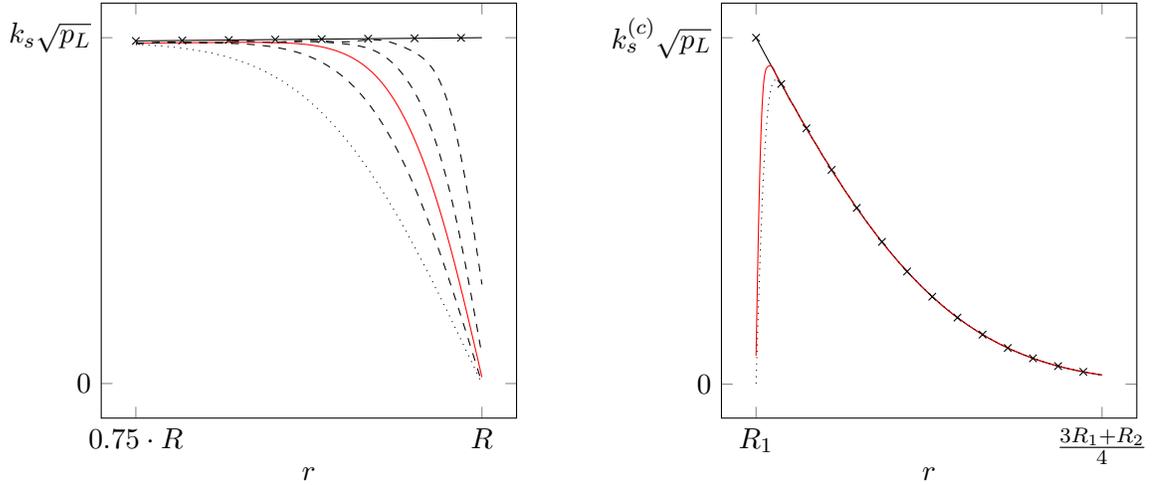
\begin{figure}[h]
	\centering
	\begin{subfigure}[t]{0.47\textwidth}
		\begin{tikzpicture}
		\begin{axis}[
		scaled ticks=false,
		width=\textwidth, 
		height=\textwidth, 
		xlabel={$r$},
		xtick={0.00225 ,0.003},
		ymin=-0.1, ymax=1.1,
		ytick={0.0,1.0},
		xticklabels={$0.75 \cdot R$ , $R$},
		yticklabels={$0$ , $k_s \sqrt{p_L}$},
		]
		\addplot[color=black,mark=x,mark repeat=20] table [x=a, y=b, col sep=comma] {data0_s_pump};\label{p20}
		\addplot[color=black,dashed] table [x=a, y=b, col sep=comma] {data1_s_pump};\label{p21}
		\addplot[color=black,dashed] table[x=a, y=b, col sep=comma] {data2_s_pump};\label{p22}
		\addplot[color=red] table [x=a, y=b, col sep=comma] {data3_s_pump};\label{p23}
		\addplot[color=black,dashed] table [x=a, y=b, col sep=comma] {data4_s_pump};\label{p24}
		\addplot[color=black,dotted] table[x=a, y=b, col sep=comma] {data5_s_pump};\label{p25}
		\end{axis}
		\end{tikzpicture}
		\caption{Solution $c_2$ for $z= 0$ in the first time interval.} \label{pic_spec_sol2}
	\end{subfigure}
	\hfill
	\begin{subfigure}[t]{0.47\textwidth}
		\begin{tikzpicture}
		\begin{axis}[
		scaled ticks=false,
		width=\textwidth, 
		height=\textwidth, 
		xtick={0.01, 0.0125},
		ytick={0.0, 1.0},
		xlabel={$r$},
		xticklabels={$R_1$, $\frac{3R_1+R_2}{4}$},
		yticklabels={$0$ , $k_s^{(c)} \sqrt{p_L}$},
		]
		\addplot[color=black,mark=x,mark repeat=40] table [x=a, y=b, col sep=comma] {data0_v_pump};
		\addplot[color=black,red] table [x=a, y=b, col sep=comma] {data2_v_pump};\\
		\addplot[color=black,dotted] table [x=a, y=b, col sep=comma] {data4_v_pump};
		\end{axis}
		\end{tikzpicture}
		\caption{Solution $u_2$ in the first time interval.} \label{pic_ves_sol2}
	\end{subfigure}
	\caption{Solutions in the second time interval as given by the Equations \eqref{t_2_g_sol} and \eqref{t_2_h_sol}. Each graph belongs to one defined time and (\ref{p20}, $t=t_1=800 \, \text{s}$) and (\ref{p25}, $t=805 \, \text{s}$). The graphs (\ref{p22}) belong to times $800\, \text{s}<t<805\, \text{s}$. The line (\ref{p23}) corresponts to $t = 801.6 \,\text{s}$. At $t=801.6$ the remaining pressure of the gaseous phase is $1.0\,\text{Pa}$. We used the same constants as in Figure \ref{fig_t_1}. In addition we used $\sigma = 5.0 \, \text{Hz}$. } \label{fig_t_2}
\end{figure}
The corresponding solutions $c_1$ and $u_1$ are plotted in Figure \ref{fig_t_2}. Again it is simple to recover the homogeneous part using the solution in the previous time interval. Thus,
\begin{align}
&\langle \varphi_{mn},g_2|_{t=t_2} \rangle =-k_s\sqrt{p_L} \langle \varphi_{mn},1 \rangle\left(\exp[-\lambda_{m,n}^2 D_s t_2]-\frac{\sigma }{2\lambda_{m,n}^2 D_s - \sigma } 
\nu_{m,n}(t_2-t_1)\right) ,\\
&\langle \psi_{n},h_2|_{t=t_2} \rangle = -k_s^{(c)}\sqrt{p_L} \langle  \psi_n,\omega  \rangle \left(\exp[-y_n^2 D_s t_2]-\frac{\sigma \kappa_{n}(t_2-t_1)}{\left(2y_n^2 D_c - \sigma\right)  }\right).
\end{align}
Using the ansatz in Equation \eqref{sed_an} one computes the remaining integrals on the right hand side of Equations \eqref{duhamel_1} and \eqref{duhamel_2} for $i=3$ as
\begin{align}
&\int_0^t \exp[-\lambda_{m,n}^2 D_s (t-s)] \langle\varphi_{m,n}, f_{3,s} \rangle ds \\
&=-k_s \sqrt{p_L}\langle\varphi_{m,n}, 1\rangle \frac{\beta\left(\sqrt{p_f}-\sqrt{p_0}\right)}{\sqrt{p_L}(1-\exp[-\beta \tau])(\lambda_{m,n}^2 D_s - \beta)} \left(\exp[-\beta t]-\exp[-\lambda_{m,n}^2D_s t] \right), \nonumber \\
&\int_0^t \exp[-y_n^2 D_c (t-s)]  \frac{\langle\psi_n, \omega f_{3,v}  \rangle}{\abs{\langle  \psi_n, \psi_n  \rangle}} ds  \\
&= \frac{-k_s^{(c)} \sqrt{p_L} \langle\psi_n, \omega   \rangle }{\abs{\langle  \psi_n, \psi_n  \rangle}} \frac{\beta \left(\sqrt{p_f}-\sqrt{p_0}\right)}{\sqrt{p_L}(1-\exp[-\beta \tau])(y_n^2 D_c - \beta)} \left(\exp[-\beta t]-\exp[-y_n^2D_c t] \right). \nonumber 
\end{align}
It is convenient to introduce the time-dependent parts of the solutions in the third time interval as
\begin{align}
\mathcal{Q}^{(s)}_{m,n}(t) \coloneqq &\exp[-\lambda_{m,n}^2 D_s (t+t_2)] -\frac{\sigma\exp[-\lambda_{m,n}^2 D_s t] }{2\lambda_{m,n}^2 D_s - \sigma } 
\nu_{m,n}(t_2-t_1)  \\
&+\frac{\beta \left(\sqrt{p_f}-\sqrt{p_0}\right)}{\sqrt{p_L}(1-\exp[-\beta \tau])(\lambda_{m,n}^2 D_s - \beta)} \left(\exp[-\beta t]-\exp[-\lambda_{m,n}^2D_s t] \right) \nonumber
\end{align}
and
\begin{align}
\mathcal{Q}^{(c)}_{n}(t) \coloneqq &\exp[-y_n^2 D_s (t+t_2)]-\frac{\sigma\exp[-y_n^2 D_s t)] }{\left(2y_n^2 D_c - \sigma\right)  }\kappa_{n}(t_2-t_1)\\
& +\frac{\beta \left(\sqrt{p_f}-\sqrt{p_0}\right)}{\sqrt{p_L}(1-\exp[-\beta \tau])(y_n^2 D_c - \beta)} \left(\exp[-\beta t]-\exp[-y_n^2D_c t] \right). \nonumber
\end{align}
Note that we shifted the time such that $\mathcal{Q}^{(c)}_{n}(t):[0,t_3-t_2] \to \R_+$ and $\mathcal{Q}^{(s)}_{m,n}(t):[0,t_3-t_2] \to \R_+$. We are finally able to state the solution in the last time interval as
	\begin{align}
	g_3(r,z,t+t_2) &= \sum_{m,n}(-1)^{n+1}\frac{8k_s \sqrt{p_L}}{x_m \mathcal{J}_1(x_m) (2n+1) \pi} \mathcal{Q}^{(s)}_{m,n}(t) \mathcal{J}_0\left( x_m \frac{r}{R} \right) \cos\left(\frac{(2n+1)\pi}{L}z\right), \label{g3_simp}\\
	h_3(r,t+t_2)  &= \sum_{n}\frac{-k_s^{(c)}\sqrt{p_L} \langle  \psi_n,\omega  \rangle}{\abs{\langle  \psi_n, \psi_n  \rangle}}   \mathcal{Q}^{(c)}_{n}(t)\left(\frac{-Y_0(y_n R_1)}{\mathcal{J}_0(y_n R_1)} \mathcal{J}_0(y_n r) + Y_0(y_n r)\right). \label{h3_simp}
	\end{align}
	The corresponding solutions $c_3$ and $u_3$ are plotted in Figure \ref{fig_t_3}.
	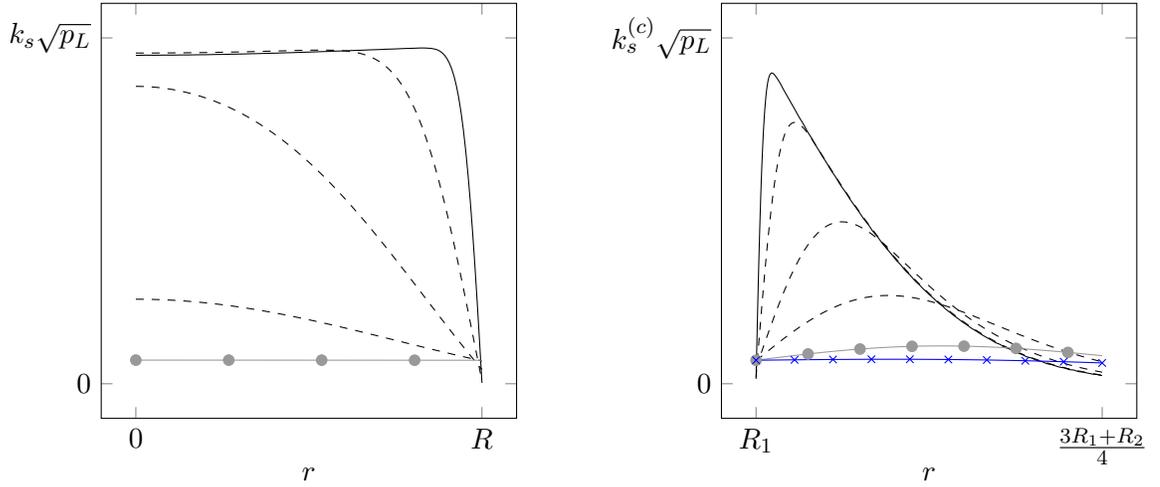
\begin{figure}[h]
		\centering
		\begin{subfigure}[t]{0.47\textwidth}
			\begin{tikzpicture}
			\begin{axis}[
			scaled ticks=false,
			width=\textwidth, 
			height=\textwidth, 
			ymin=-0.1, ymax=1.1,
			xlabel={$r$},
			xtick={0.0 ,0.003},
			ytick={0.0,1.0},
			xticklabels={$0$ , $R$},
			yticklabels={$0$ , $k_s \sqrt{p_L}$},
			]
			\addplot[color=black] table [x=a, y=b, col sep=comma] {data0_s_rel};\label{p30}
			\addplot[color=black,dashed] table [x=a, y=b, col sep=comma] {data1_s_rel};\label{p31}
			\addplot[color=black,dashed] table[x=a, y=b, col sep=comma] {data2_s_rel};\label{p32}
			\addplot[color=black,dashed] table [x=a, y=b, col sep=comma] {data3_s_rel};\label{p33}
			\addplot[color=black!40!white,mark=*,mark repeat=40] table [x=a, y=b, col sep=comma] {data4_s_rel};\label{p34}
			\end{axis}
			\end{tikzpicture}
			\caption{Solution $c_3$ for $z=0$ in the first time interval.} \label{pic_spec_sol3}
		\end{subfigure}
		\hfill
		\begin{subfigure}[t]{0.47\textwidth}
			\begin{tikzpicture}
			\begin{axis}[
			scaled ticks=false,
			width=\textwidth, 
			height=\textwidth, 
			xtick={0.01, 0.0125},
			ytick={0.0, 1.0},
			ymin=-0.1, ymax=1.1,
			xlabel={$r$},
			xticklabels={$R_1$, $\frac{3R_1+R_2}{4}$},
			yticklabels={$0$ , $k_s^{(c)} \sqrt{p_L}$},
			]
			\addplot[color=black] table [x=a, y=b, col sep=comma] {data0_v_rel};
			\addplot[color=black,dashed] table [x=a, y=b, col sep=comma] {data1_v_rel};
			\addplot[color=black,dashed] table [x=a, y=b, col sep=comma] {data2_v_rel};
			\addplot[color=black,dashed] table [x=a, y=b, col sep=comma] {data3_v_rel};
			\addplot[mark=*,color=black!40!white,mark repeat=120] table [x=a, y=b, col sep=comma] {data4_v_rel};
			\addplot[color=blue,mark=x] table [x=a, y=b, col sep=comma] {data7_v_rel};\label{blue_line}
			\end{axis}
			\end{tikzpicture}
			\caption{Solution $u_3$ in the first time interval.} \label{pic_ves_sol3}
		\end{subfigure}
		\caption{Solutions in the third time interval as given by the Equations \eqref{g3_simp} and \eqref{h3_simp}. Each graph belongs to one defined time and (\ref{p30}, $t=t_2=801.6 \, \text{s}$) and (\ref{p34}, $t=2401.6 \, \text{s}$). The graphs (\ref{p32}) belong to times $801.6\, \text{s}<t<2401.6\, \text{s}$. We used the same constants as in Figure \ref{fig_t_2}. In addition we used $\beta =9.0e^{-3}$, $\tau=1500 \, \text{s}$, $p_f = 13.85\,\text{Pa}$. Note that for (\ref{blue_line}) at $t-t_2=3600 \, \text{s}$ the gradient vanishes. }\label{fig_t_3}
	\end{figure}
	\subsection{Pressure in the Gaseous Phase}
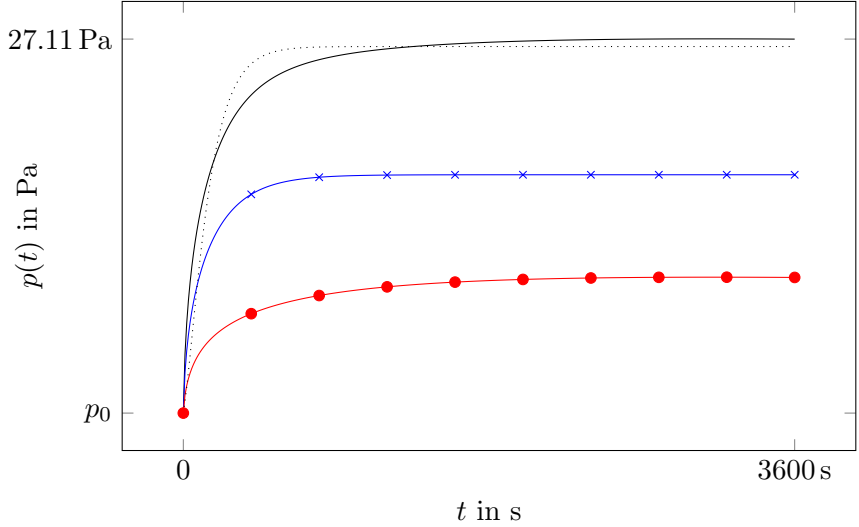
\begin{figure}[!htb]
		\centering
		\begin{tikzpicture}
		\begin{axis}[
		scaled ticks=false,
		width=0.75\textwidth, 
		height=0.5\textwidth, 
		xlabel={$t$ in s},
		ylabel={$p(t) $ in Pa},
		xtick={0.0 ,3600},
		ytick={1.0,27.11},
		xticklabels={$0$ , $3600 \, \text{s}$},
		yticklabels={$p_0$ , $27.11 \, $Pa},
		]
		\addplot[color=black] table [x=a, y=b, col sep=comma] {data_p_combined};\label{pf}
		\addplot[color=blue,mark repeat=40,mark=x] table [x=a, y=b, col sep=comma] {data_p_s};\label{ps}
		\addplot[color=red,mark repeat=40,mark=*] table[x=a, y=b, col sep=comma] {data_p_v};\label{pv}
		\addplot[color=black,dotted] table[x=a, y=b, col sep=comma] {ansatz};\label{ansatz}
		\end{axis}
		\end{tikzpicture}\caption{Pressure increase in the last time interval as given in Equation \eqref{final_pressure}. We assumed the same parameters as in Figure \ref{fig_t_3} and $T=673.15 \, \text{K}$ and  $L_{out} = 0.08 \, \text{m}$. Here (\ref{pf}) corresponds to $p_{H_2}(t) + p_0$ and (\ref{ansatz}) corresponds to the ansatz $p_3(t+t_2)$ chosen in Equation \eqref{sed_an}. The graph (\ref{ps}) corresponds to the pressure increase caused only by the specimen. The graph (\ref{pv}) corresponds to the pressure increase caused only by the container as given in the Equations \eqref{pressure_spec} and \eqref{pressure_container}.} \label{fig_rel_main}
	\end{figure}

	Since the experiment is kept at a constant temperature $T$ one can apply the ideal gas law
	\begin{align} \label{final_pres}
	p_{H_2}(t) = \frac{ \left(n_{H}(t) + n_{H}^{(c)}(t)  \right)\textit{\textbf{R}} T}{ 2 \left( V_{App} - V_{s} \right) }.
	\end{align}
	Here $n_{H}(t)$ and $n_{H}^{(c)}(t)$ denotes the molar amount of mono-atomic hydrogen released from the surfaces of the specimen and the inner surface of the container, $V_{App}$ is the volume enclosed by the container and $V_s$ is the volume of the specimen. Since the hydrogen recombines for diatomic hydrogen in the gaseous phase, one has to take a factor two into account. The gas constant is denoted with $\textit{\textbf{R}}$.\\
	Using the Equations \eqref{g3_simp} and \eqref{h3_simp} one can find an expression for the flux of desorbed hydrogen as
	\begin{align} \label{Flux}
	\frac{\dot{J}(t)}{2 \pi} = &\underset{\eqqcolon \dot{J}_1(t)}{\underbrace{-D_s R \int_{-\frac{L}{2}}^{\frac{L}{2}} \partial_r g_3(R,z,t+t_2) dz}} \quad \underset{\eqqcolon \dot{J}_2(t)}{\underbrace{-2 D_s \int_{0}^{R} \partial_z g_3\left(r,\frac{L}{2},t+t_2\right) r dr}}  \\\nonumber
	&\underset{\eqqcolon \dot{J}_3(t)}{\underbrace{-D_{v}R_1 \int_{-\frac{L_{out}}{2}}^{\frac{L_{out}}{2}} \partial_r u_{3}(R_1,t+t_2) dz }}.
	\end{align}
	Here $L_{out}$ denotes the length of the container. Note that we assumed $L_{out}$ to be infinitely large. This assumption is still good if $\frac{L_{out}}{R_1} \gg 1$. One could extend this model for finitely long containers as it was done for the specimen. Note that there is an additional contribution to the flux from the caps of the container, that we didn't take into account. By the same argument $\frac{L_{out}}{R_1} \gg 1$ we assume, the contribution of the caps to be small compared to the contribution of the inner coat. This simplification was done, since it is not clear which boundary conditions should be assumed at the connection of the caps of the container and its coat.\\
	Note that since $\partial_r g_3 = \partial_r c_3$, one can actually use the function $g_3$ in Equation \eqref{Flux} instead of $c_3$ but for the concentration in the container $u_3$ one has to use the explicit relation in Equation \eqref{cast_u_h}. One easily checks
	\begin{align}
	\partial_r u_3(R_1,t+t_2) = \partial_r h_3(R_1,t+t_2) + \frac{k_s^{(c)} \sqrt{p_3(t+t_2)}}{R_1 \log\left(\frac{R_1}{R_2}\right)}. \label{dru}
	\end{align}
	Note that the second term on the right hand side in Equation \eqref{dru} does not vanish for $t\to \infty$, since the ansatz in Equation \eqref{sed_an} is an increasing function. As one can check in Figure \ref{pic_ves_sol3} the gradient of the concentration at $r=R_1$ is positive for reasonable times but will flip sign at some point (at $t-t_2\approx 3600 \, \text{s}$). Therefore some hydrogen will be solved in the container for $t-t_2 > 3600 \, \text{s}$ and the pressure in the gaseous phase will then begin to drop. This is not correctly described with the chosen ansatz. In order to describe this one could extend the ansatz by replacing $p_3(t)$ with $p_3(t) \chi(t)$, where
	\begin{align}
	\chi(t) \coloneqq \frac{1}{1+\exp[-(t-\xi)q]}
	\end{align}\\
	with the free parameters $\xi$ and $q$. We restrict the evaluation to times $t-t_2 \leq 3600 \, \text{s}$ and the given ansatz in Equation \eqref{sed_an}. \\
	It is tedious but easy to compute the currents $\dot{J}_k$ for $k \in \{1,2,3\}$ in Equation \eqref{Flux}. They are
	\begin{align}
	\dot{J}_1(t) &= L \sum_{m,n} \frac{16 k_s D_s \sqrt{p_L}}{(2n+1)^2\pi^2}  \mathcal{Q}^{(s)}_{m,n}(t),\\
	\dot{J}_2(t) &= \frac{R^2}{L} \sum_{m,n} \frac{16 k_s  D_s \sqrt{p_L}}{ x_m^2} \mathcal{Q}^{(s)}_{m,n}(t), \\
	\dot{J}_3(t) &=L_{out} R_1\sum_{n}\left[\frac{k_s^{(c)} D_c\sqrt{p_L} \langle  \psi_n,\omega  \rangle y_n}{\abs{\langle  \psi_n, \psi_n  \rangle}} \left(\frac{Y_0(y_n R_1)}{\mathcal{J}_0(y_n R_1)} \mathcal{J}_1(y_n R_1) - Y_1(y_n R_1)\right)\mathcal{Q}^{(c)}_{n}(t)\right] \\
	& -  L_{out}\frac{k_s^{(c)} D_c \sqrt{p_3(t+t_2)}}{ \log\left(\frac{R_1}{R_2}\right)}. \nonumber
	\end{align}
	Note that for $R\ll L$ the current $\dot{J}_2$ is indeed insignificant. The simplification $\dot{J}_2 \equiv 0$ does not simply recover the results in \cite{SL1999} since the eigenvalues in Equation \eqref{Eigenval_s} depend on $L$. The same arguments also hold for the container. Note the units  $\left[\mathcal{Q}^{(c)}_{n}\right] = \left[ \mathcal{Q}^{(s)}_{m,n} \right]=1$ and $\left[y_n\right]=\text{m}^{-1}$, such that indeed $\left[\dot{J}_k\right] = \frac{\text{mol}}{\text{s}}$.\\
	Integrating these molecular fluxes with respect to the time gives the molar amounts  $n_{H}(t)$ and $n_{H}^{(c)}(t)$  in Equation \eqref{final_pres} up to a factor of $2\pi$. We finally derive
	\begin{align} \label{final_pressure}
	\frac{p_{H_2}(t)}{2\pi}= &\frac{ \textit{\textbf{R}} T  }{ 2 \left( V_{App} - V_{s} \right) } \Bigg\{ \sum_{m,n}16 k_s  D_s \sqrt{p_L}\left( \frac{L}{(2n+1)^2\pi^2} + \frac{R^2}{L x_m^2} \right) \int_{0}^{t}\mathcal{Q}^{(s)}_{m,n}(t') dt' \\ \nonumber
	&+L_{out} R_1\sum_{n}\frac{k_s^{(c)} D_c\sqrt{p_L} \langle  \psi_n,\omega  \rangle y_n}{\abs{\langle  \psi_n, \psi_n  \rangle}} \left(\frac{Y_0(y_n R_1)}{\mathcal{J}_0(y_n R_1)} \mathcal{J}_1(y_n R_1) - Y_1(y_n R_1)\right)\int_{0}^{t}\mathcal{Q}^{(c)}_{n}(t')dt' \\\nonumber
	& -  L_{out}\frac{k_s^{(c)} D_c }{ \log\left(\frac{R_1}{R_2}\right)} \int_{0}^{t} \sqrt{p_3(t'+t_2)} dt' \Bigg\}.
	\end{align}
	Since we shifted the time in the Definition of $\mathcal{Q}^{(s)}_{m,n}$ and $\mathcal{Q}^{(c)}_{n}$ one has $p_{H_2} : [0,t_3-t_2] \to \R$. The integrals in Equation \eqref{final_pressure} are easy to solve.\\
	If the ansatz in Equation \eqref{sed_an} is indeed the solution to the problem one would have $p_3(t+t_2) =p_{H_2}(t)$ for ($p_f$,$\tau$,$\beta$) chosen correctly. Minimizing 
	\begin{align}
	\varepsilon(p_f,\tau,\beta) = \int_{0}^{t_3-t_2} \abs{p_{H_2}(t)-p_3(t+t_2)} dt
	\end{align}
	gives the best approximation to the solution. It may be convenient to fix $\tau = t_3-t_2$ and minimize with respect to $(p_f,\beta)$, since we expect the minimizer not to be unique.\\
	The resulting release graph is plotted in Figure \ref{fig_rel_main}. We used the same paramters as for Figure \ref{fig_t_1}, \ref{fig_t_2} and \ref{fig_t_3}. We assumed $L_{out} = 0.08 \, \text{m}$.
	\subsection{Comparsion with Non-Interacting Surfaces} \label{comparsion}
	We compare now the resulting behavior in Figure \ref{fig_rel_main} with the case, of non-interacting specimen and container.\\
	For this we define
	\begin{align}
	&\frac{p_s(t)}{2 \pi} \coloneqq \frac{ \textit{\textbf{R}} T  }{ 2 \left( V_{App} - V_{s} \right) } \sum_{m,n}16 k_s  D_s \sqrt{p_L}\left( \frac{L}{(2n+1)^2\pi^2} + \frac{R^2}{L x_m^2} \right) \int_{0}^{t}\mathcal{Q}^{(s)}_{m,n}(t') dt', \label{pressure_spec} \\
	&\frac{p_v(t)}{2 \pi}\coloneqq \frac{ \textit{\textbf{R}} T  }{ 2 \left( V_{App} - V_{s} \right) } \Bigg\{ \label{pressure_container} \\ \nonumber
	& L_{out} R_1\sum_{n}\frac{k_s^{(c)} D_c\sqrt{p_L} \langle  \psi_n,\omega  \rangle y_n}{\abs{\langle  \psi_n, \psi_n  \rangle}} \left(\frac{Y_0(y_n R_1)}{\mathcal{J}_0(y_n R_1)} \mathcal{J}_1(y_n R_1) - Y_1(y_n R_1)\right)\int_{0}^{t}\mathcal{Q}^{(c)}_{n}(t')dt' \\\nonumber
	& -  L_{out}\frac{k_s^{(c)} D_c }{ \log\left(\frac{R_1}{R_2}\right)} \int_{0}^{t} \sqrt{p_3(t'+t_2)} dt' \Bigg\}.
	\end{align}
	\begin{figure}[!htb]
		\centering
		\begin{tikzpicture}
		\begin{axis}[
		scaled ticks=false,
		width=0.75\textwidth, 
		height=0.5\textwidth, 
		xlabel={$t$ in s},
		ylabel={$p(t) $ in Pa},
		xtick={0.0 ,3600},
		ytick={1.0,28.79},
		xticklabels={$0$ , $3600 \, \text{s}$},
		yticklabels={$p_0$ , $28.79$ Pa},
		]
		\addplot[color=black, dashed] table [x=a, y=b, col sep=tab] {data_p_combined_sedano};\label{pf3}
		\addplot[color=black] table [x=a, y=b, col sep=comma] {data_p_combined};\label{pf2}
		\addplot[color=blue,dashed] table [x=a, y=b, col sep=comma] {data_p_s_sedano};\label{ps_sedano}
		\addplot[color=blue,mark repeat=40,mark=x] table [x=a, y=b, col sep=comma] {data_p_s};\label{ps_2}
		\addplot[color=red,dashed] table[x=a, y=b, col sep=comma] {data_p_v_sedano};\label{pv_sedano}
		\addplot[color=red,mark repeat=40,mark=*] table[x=a, y=b, col sep=comma] {data_p_v};\label{pv_2}	
		\addplot[color=blue,dotted] table[x=a, y=b, col sep=comma] {ansatz_s};\label{ansatz_s}
		\addplot[color=black,dotted] table[x=a, y=b, col sep=comma] {ansatz_c};\label{ansatz_v}
		\end{axis}
		\end{tikzpicture}\caption{Pressure increase in the gas release phase. We assumed the same parameters as in Figure \ref{fig_t_3} and $T=673.15 \, \text{K}$ and  $L_{out} = 0.08 \, \text{m}$. We fit the ansatz in Equation \eqref{sed_an} to the specimen and container independently. The lines (\ref{ansatz}) correspond to the fitted ansatz. The pressure increase (up to $p_0$) of the specimen corresponds to (\ref{ps_sedano}) and for the container to (\ref{pv_sedano}). Here (\ref{pf3}) is simply  the sum of (\ref{ps_sedano}) and (\ref{pv_sedano}) (minus $p_0$). The lines (\ref{ps_2}) and (\ref{pv_2}) are the same as in Figure \ref{fig_rel_main}.} \label{fig_rel_ind}
	\end{figure}
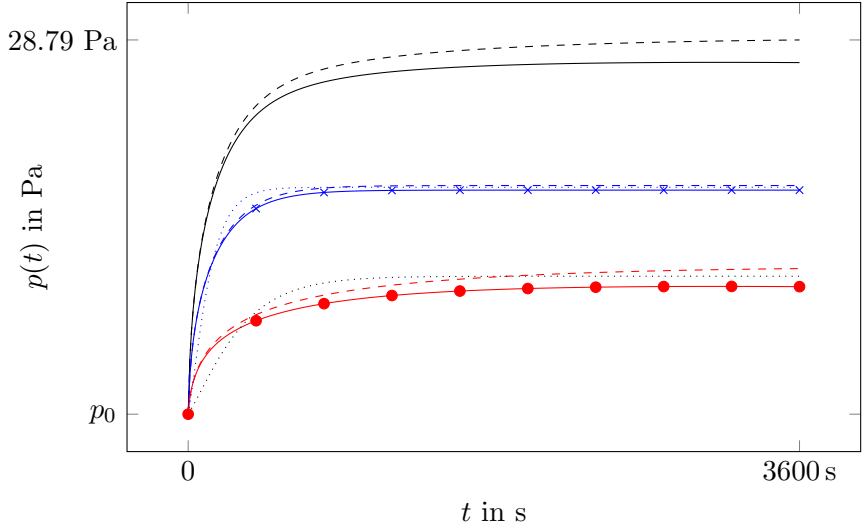 
	In contrary to the previous case we neglect the container and fit directly the ansatz in Equation \eqref{sed_an} to the pressure increase in Equation \eqref{pressure_spec}. Independently we fit the ansatz to Equation \eqref{pressure_container} recovering the pressure increase in the absence of the specimen (We still assume that the gaseous phase inhabits the same volume). This is the same as measuring the pressure increase in a zero experiment without specimen. Both resulting graphs are plotted in Figure \ref{fig_rel_ind}.\\
	One recognizes that one overestimates the pressure increase of the container in Figure \ref{fig_rel_ind}, since the specimen and the container desorbes hydrogen at the same time and hence both surfaces contribute less to the combined pressure increase.\\
	Comparing both maximal values in Figure \ref{fig_rel_main} and \ref{fig_rel_ind} shows that both values differ by approximately $ 6.2\, $\%.
	\subsection{Comparison with Numerical Simulations}
	\begin{figure}[!htb]
		\begin{center}
			\includegraphics[width=0.9\textwidth]{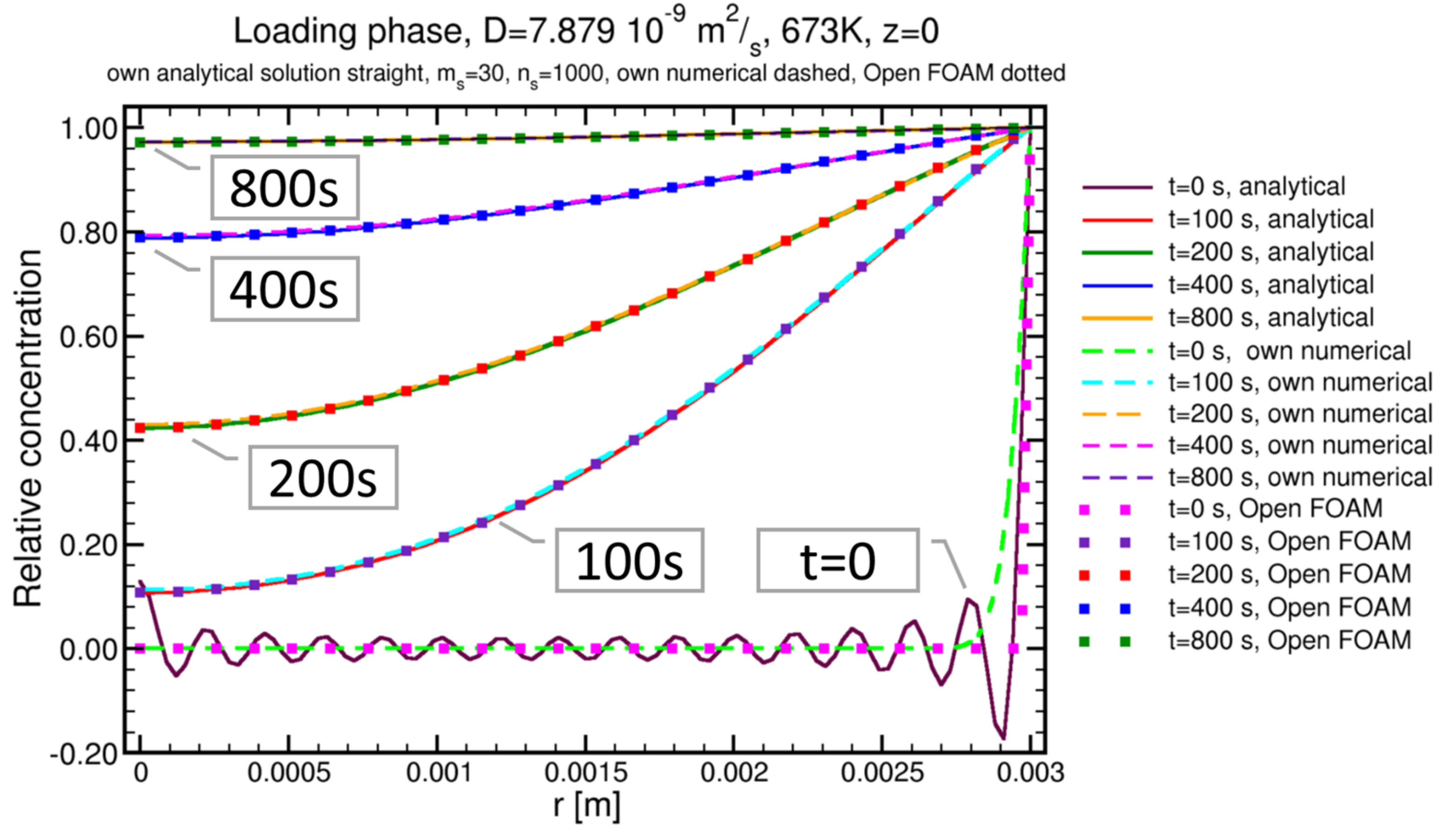}
			\caption{Concentration distribution at $z\equiv 0$ for different times in the first time interval. The numbers $\text{m}_s$ and $\text{n}_s$ denote the amount of zeros of Bessel's Functions and the amount of zeros of the polynomial in Equation \eqref{polynom} used. \label{fred2}}
		\end{center}
	\end{figure}
\begin{figure}[!htb]
		\begin{center}
			\includegraphics[width=0.9\textwidth]{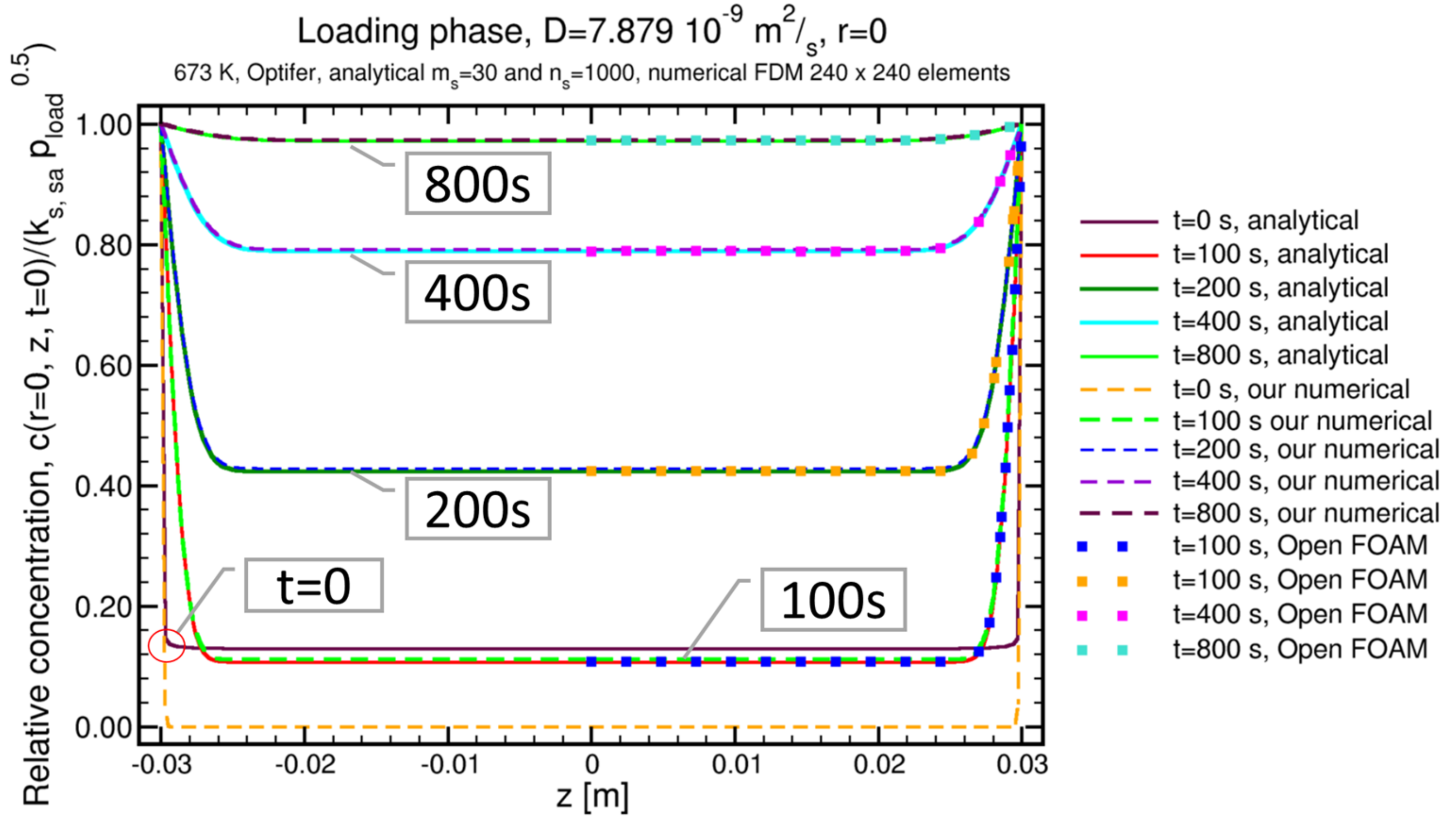}
			\caption{Concentration distribution at $r\equiv 0$ for different times in the first time interval. The numbers $\text{m}_s$ and $\text{n}_s$ denote the amount of zeros of Bessel's Functions and the amount of zeros of the polynomial in Equation \eqref{polynom} used. Note that the deviation at $t=0$ and compare this to Figure \ref{fred2}.\label{fred1}}
		\end{center}
	\end{figure}

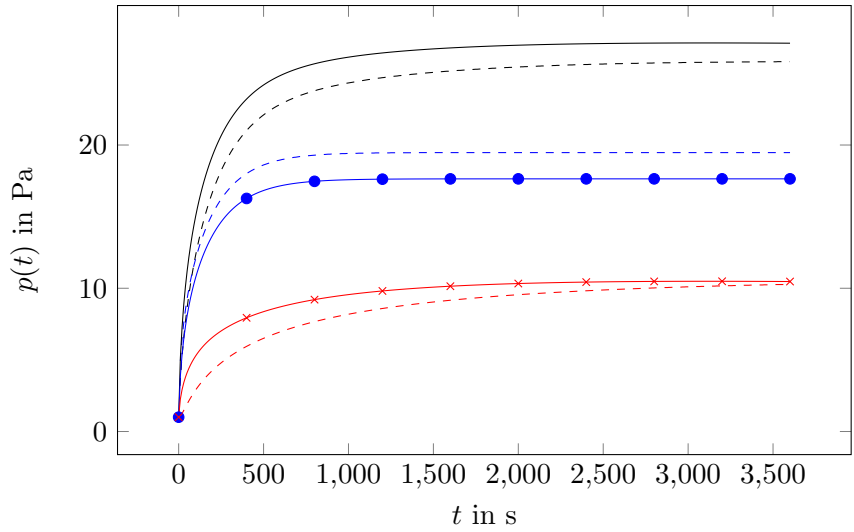
\begin{figure}[!htb]
		\centering
		\begin{tikzpicture}
		\begin{axis}[
		scaled ticks=false,
		width=0.75\textwidth, 
		height=0.5\textwidth, 
		xlabel={$t$ in s},
		ylabel={$p(t)$ in Pa},
		]
		\addplot[color=black, dashed] table [x=a, y=b, col sep=tab] {specconffdm.dat};\label{weth1}
		\addplot[color=blue, dashed] table [x=a, y=b, col sep=tab] {specfdm.dat};\label{weth2}
		\addplot[color=red, dashed] table [x=a, y=b, col sep=tab] {conffdm.dat};\label{weth3}
		\addplot[color=black] table [x=a, y=b, col sep=comma] {data_p_combined};\label{an1}
		\addplot[color=blue,mark repeat=40,mark=*] table [x=a, y=b, col sep=comma] {data_p_s};\label{an2}
		\addplot[color=red,mark repeat=40,mark=x] table[x=a, y=b, col sep=comma] {data_p_v};\label{an3}
		\end{axis}
		\end{tikzpicture}\caption{Comparsion between the derived analytical model assigned with solid lines and some FDM-Solver assigned with dashed lines. Note that (\ref{an1}) resembles the pressure increase in the gaseous phase, (\ref{an2}) is the contribution of the specimen and (\ref{an3}) is the contribution of the container.} \label{vgl_fdm}
	\end{figure}

\noindent
	Instead of the analytical approach using Duhamel's formula one can use some finite difference method (FDM) \cite{VDWA2019} or finite volume method (FVM) \cite{VP2018}. We compare the numerical result using (FDM) and Open FOAM with the analytical solution in Figure \ref{fred1} and \ref{fred2}. The solution using Duhamel's formula indeed recovers the numerical results at least for any positive time $t>0$ in the first time interval. The oscillating behavior at $t=0$ is no surprise, since one tries to approximate a non-continous function with Bessel's functions.
	\noindent
	In the second and especially in the third time interval these numerical approaches are time consuming. In Figure \ref{vgl_fdm} we compared the resulting release graphs of both methods. One notes that the FDM-Solver struggles describing the pressure increase caused by the container for small times. The FDM-Solver predicts a smaller pressure increase.
	 \noindent
	\subsection{Example Application} \label{choice}
	To obtain the figures in the paper on hands we used the parameters stated beneath the Figures \ref{fig_t_1}-\ref{fig_t_3}. Especially we used for the specimen ($k_s=1.829e^{-3}\frac{\text{mol}}{\text{m}^3 \sqrt{\text{Pa}}}$, $D_s=7.879e^{-9} \, \frac{\text{m}^2}{\text{s}}$) and for the container ($k_s^{(c)}=5.914e^{-4}\frac{\text{mol}}{\text{m}^3 \sqrt{\text{Pa}}}$, $D_c= 8.257e^{-10}\, \frac{\text{m}^2}{\text{s}}$). These parameters resemble a specimen made of steel (Eurofer, $9\%_{\text{wt}}$ Cr, $1\%_{\text{wt}}$ W) and a container made of copper. The details on what materials are preferable for the specific experiment are discussed in \cite{VDWA2019}.\\
	In view of Figure \ref{fig_rel_main} it is conceivable to use a copper container since the signal of desorbed hydrogen from the specimen still dominates the signal of the container. In Figure \ref{molar_amount} one recognizes that the amount of hydrogen stored in the container is indeed higher than the amount stored in the specimen. In order to keep the amount of hydrogen in the container as small as possible one could reduce the charging time $t_1$.
	\begin{figure}[!htb]
		\centering
		\begin{tikzpicture}
		\begin{axis}[
		scaled ticks=false,
		width=0.75\textwidth, 
		height=0.5\textwidth, 
		xlabel={$t$ in s},
		ylabel={$n(t)$ in mol},
		]
		\addplot[color=blue,only marks,mark=x] table [x=a, y=b, col sep=tab] {consumption1};\label{c1}
		\addplot[color=red!40!white,only marks,mark=o] table[x=a, y=b, col sep=tab] {consumption2};\label{c2}
		{ansatz};
		\end{axis}
		\end{tikzpicture}\caption{The molar amount of hydrogen stored in the specimen (\ref{c1}) and in the container (\ref{c2}).} \label{molar_amount}
	\end{figure}
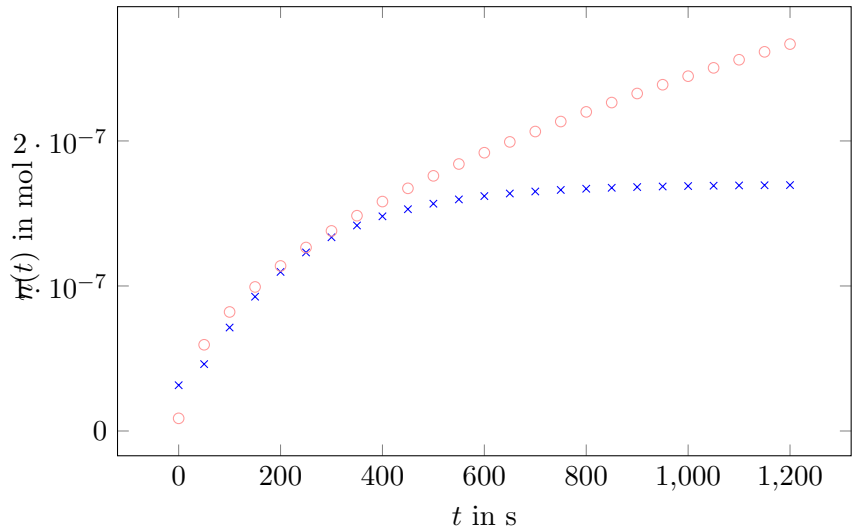 \noindent
	
	\section{CONCLUSIONS}
	Since transport parameters cannot be determined directly one fits models depending on these parameters to actual data of an experiment using some branch-and-bound algorithm. These obtained parameters such as Sievert's constant and diffusivity fit as good as possible in the given model. This underlines the fact, that these obtained parameters depend on the models used as we explained in Section \ref{comparsion}. In order to find the best possible model we derived solutions to the Diffusion Equation, that feature:
	\begin{enumerate}
		\item \textbf{Fast: }Evaluating the Equation \eqref{final_pressure} at $350$ data points takes less than $7.5 \, \text{s}$  (Core: Intel i5 of the 8th. generation). Note, that the code used is not optimized for speed and is coded in Python.   
		\item \textbf{Time-Dependent BC.: }the transformations in the Equations \eqref{transform_1} and \eqref{cast_u_h} and Duhamel's formula guarantee that the assumed boundary conditions are fulfilled.
		\item \textbf{Interacting Surfaces: } The solution in Equation \eqref{final_pressure} takes the inner wall of the container into account such that the interaction between specimen and container is not simply neglected.
		\item \textbf{Variable Parameters: }The length of each phase of the experiment can be tuned with the paramteres $t_k$ for $k \in \{1,2,3\}$. We can simulate different parameters of $k_s$ and $D$ for the specimen and the container.
	\end{enumerate} 
	As explained in Section \ref{choice} the derived model is suitable for simulating different material constants and hence can help finding the optimal choice for the container material. We indeed showed in Section \ref{comparsion} that the interaction between container and specimen can not be neglected.\\
	We mention again, that the derived model uses the ansatz in Equation \eqref{sed_an} and hence is an approximation to the correct solution to the problem stated in Section \ref{equations}.\\
	Note that the whole problem in finding $p(t)$ can be formulated as a fixed-point problem. Solving this fixed-point problem with standard iterative techniques such as Banach's fixed-point Theorem could give the correct solution but is again computational costly.

	
	\section*{ACKNOWLEDGEMENTS}
	
	The authors are grateful for the support by Dirk Hundertmark and Ron Dagan during this interdisciplinary project. This work has been carried out within the framework of the EUROfusion Consortium, and has received funding from the Euratom research and training program 2019-2020 under grant agreement No.
	633053. The views and opinions expressed herein do not necessarily
	reflect those of the European Commission. The authors are also thanking for support and fundings by MathSEE at KIT regarding the project: Neue L\"osungen der
	Kontinuit\"atsdiffernetialgleichung mit Phasengleichgewicht zur Verbesserung der Ergebnisse bei der Auswertung von Experimenten.

	
	\appendix
	\section{}\label{coefficients} \noindent
	It remains to compute the coefficients in Equation \eqref{sol_1} and \eqref{sol_2} for the various time intervals. We start with the simplest case $i=1$.\\
	For the shifted concentration distribution in the specimen we have to compute
	\begin{align}
	-\langle  \varphi_{m,n}, k_s \sqrt{p_L} \rangle   &=  -k_s \sqrt{p_L}  \iint_{[0,R]\times[-\frac{L}{2},\frac{L}{2}]} f(r,z) r \frac{d(r,z)}{ L}.\\
	&= \frac{-2k_s \sqrt{p_L}}{R \mathcal{J}_1(x_m)} \iint_{[0,R]\times[-\frac{L}{2},\frac{L}{2}]} \mathcal{J}_0\left( x_m \frac{r}{R} \right) \cos\left(\frac{(2n+1)\pi}{L}z\right)r \frac{d(r,z)}{ L}. 
	\end{align}
	The integral over $z$ is easy to compute
	\begin{align}
	\int_{[-\frac{L}{2},\frac{L}{2}]}\cos\left(\frac{(2n+1)\pi}{L}z\right) \frac{dz}{L}=\frac{(-1)^{n}2}{(2n+1) \pi}.
	\end{align}
	Using standard techniques one obtains the radial integral as
	\begin{align}
	\int_{[0,R]}  \mathcal{J}_0\left( x_m \frac{r}{R} \right) r dr= \frac{R^2}{x_m^2}\int_0^{x_m} \mathcal{J}_0(t) tdt = \frac{R^2 \mathcal{J}_1(x_m)}{x_m}.
	\end{align}
	Note that by the series expansion of Bessel's functions \cite{SA1947}[\S\,19.34] around zero one directly concludes $\partial_r r^n \mathcal{J}_n(r) = r^n \mathcal{J}_{n-1}(r)$.\\
	Thus, the coefficients read
	\begin{align}\label{coeff_s_1}
	-\langle  \varphi_{m,n}, k_s \sqrt{p_L} \rangle =(-1)^{n+1}\frac{4Rk_s \sqrt{p_L}}{x_m (2n+1) \pi}.
	\end{align}
	Note that in this first interval $f_1(t) = -k_s \partial_t \sqrt{p_L} \equiv 0$.\\
	In the container the situation differs. Firstly one notes that for any time-interval it is necessary to compute the integral
	\begin{align} \label{normal_int_theta}
	\langle  \psi_n, \psi_n  \rangle = \left(\frac{Y_0(y_n R_1)}{\mathcal{J}_0(y_n R_1)} \right)^2 \int_{R_1}^{R_2} \mathcal{J}_0(y_n r)^2 r dr &- 2\frac{Y_0(y_n R_1)}{\mathcal{J}_0(y_n R_1)} \int_{R_1}^{R_2} \mathcal{J}_0(y_n r) Y_0(y_n r) r dr  \\
	&+  \int_{R_1}^{R_2} Y_0(y_n r)^2 rdr.  \nonumber
	\end{align}
	For $i>1$ one has to find additionaly
	\begin{align}
	\langle \psi_n, \omega \rangle = \int_{R_1}^{R_2} \left( \frac{-Y_0(y_n R_1)}{\mathcal{J}_0(y_n R_1)} \mathcal{J}_0(y_n r) + Y_0(y_n r) \right) \frac{\log(r) - \log(R_2)}{\log(R_1) - \log(R_2)} r dr. 
	\end{align}
	We shall briefly discuss the solutions here. Note that for Neumann's Functions one also has  $\partial_r r^n Y_n(r) = r^n Y_{n-1}(r)$ using the definition of $Y_n$ in terms of $\mathcal{J}_n$ given for example in \cite{SA1947}[\S\,19.33]. The integrals involving two Bessel's functions can be solved using the fact that they are solutions to the Bessel differential equation
	\begin{align}
	&\left(r \partial_r + r^2 \partial_r^2 \right)\mathcal{J}_0( r) = -r^2 \mathcal{J}_0( r) \\
	\Leftrightarrow\quad  &r\partial_r \left( r \partial_r\mathcal{J}_0( r)  \right) =  -r^2 \mathcal{J}_0( r)\\
	\Leftrightarrow\quad  &2r\partial_r \left( r \partial_r\mathcal{J}_0( r)  \right)\partial_r \mathcal{J}_0(r) =  -2r^2 \mathcal{J}_0( r) \partial_r \mathcal{J}_0(r)=-r^2\partial_r\left( \mathcal{J}_0(r)^2 \right) \\
	\Leftrightarrow\quad  & \partial_r  \left(r \partial_r \mathcal{J}_0(r) \right)^2 = -r^2\partial_r\left( \mathcal{J}_0(r)^2 \right).
	\end{align}
	Integrating both sides with respect to $r$ one concludes by integrating by parts,
	\begin{align}
	\left[ r^2 \left(\partial_r \mathcal{J}_0(r) \right)^2 \right]^b_a &= -\int_{a}^{b} r^2\partial_r\left( \mathcal{J}_0(r)^2 \right) dr \\
	&= -\left[\mathcal{J}_0(r)^2 r^2 \right]^b_a +2 \int_{a}^{b}\mathcal{J}_0(r)^2 r dr. 
	\end{align}
	Note that $\partial_r \mathcal{J}_0(r) = \mathcal{J}_{-1}(r) = -\mathcal{J}_1(r)$ and hence
	\begin{align}
	\int_{R_1}^{R_2}\mathcal{J}_0(y_n r)^2 r dr = \frac{1}{y_n^2} \left[\frac{r^2}{2}\left(\mathcal{J}_0(r)^2 + \mathcal{J}_1(r)^2\right) \right]^{y_n R_2}_{y_n R_1}.
	\end{align}
	The remaining integrals in Equation \eqref{normal_int_theta} follow similar. It remains to compute the integrals involving the logarithm as for example
	\begin{align}
	y_n^2 \int_{R_1}^{R_2} \mathcal{J}_0(y_n r) \log(r) rdr &= \int_{y_n R_1}^{y_n R_2} \mathcal{J}_0(r)r \log \left( \frac{r}{y_n} \right)  dr \\
	&=\left[ r \mathcal{J}_1(r) \log\left(\frac{r}{y_n}\right) \right]_{y_n R_1}^{y_n R_2} - \int_{y_n R_1}^{y_n R_2} \mathcal{J}_1(r) dr \\
	&=\left[ r \mathcal{J}_1(r) \log\left(\frac{r}{y_n}\right) + \mathcal{J}_0(r) \right]_{y_n R_1}^{y_n R_2}. 
	\end{align}
	The integral involving $Y_0$ and the logarithm follows similar. We summarize
	\begin{align}
	\langle  \varphi_{m,n}, \varphi_{m,n} \rangle&=1, \\
	\langle  \varphi_{m,n}, 1 \rangle&=(-1)^{n}\frac{4 R}{x_m (2n+1) \pi}, \\ \label{t_1}
	\langle  \psi_n, 1 \rangle &= \frac{-Y_0(y_n R_1)}{\mathcal{J}_0(y_n R_1)y_n}\left( R_2 \mathcal{J}_1(R_2) - R_1 \mathcal{J}_1(R_1)\right) + \frac{1}{y_n}\left( R_2 Y_1(R_2) - R_1Y_1(R_1) \right), \\ \label{t_2}
	\langle  \psi_n, \psi_n \rangle &= \left( \frac{Y_0(y_n R_1)}{\mathcal{J}_0(y_n R_1)} \right)^2 \left[\frac{r^2}{2 y_n^2}\left(\mathcal{J}_0(r)^2 + \mathcal{J}_1(r)^2\right) \right]^{y_n R_2}_{y_n R_1} \\ \nonumber
	&+ \left[\frac{r^2}{2 y_n^2} \left(Y_0(r)^2 + Y_1(r)^2\right) \right]^{y_n R_2}_{y_n R_1}\\
	&-2 \frac{Y_0(y_n R_1)}{\mathcal{J}_0(y_n R_1)}  \left[\frac{r^2}{2 y_n^2} \left(\mathcal{J}_0(r)Y_0(r) + \mathcal{J}_1(r)Y_1(r)\right) \right]^{y_n R_2}_{y_n R_1},\nonumber\\ \label{t_3}
	\langle  \psi_n, \omega \rangle &= \frac{-Y_0(y_n R_1)}{\mathcal{J}_0(y_n R_1) \log(\frac{R_1}{R_2})y_n^2} \left[r \mathcal{J}_1(r) \log\left(\frac{r}{y_n}\right) + \mathcal{J}_0(r)\right]^{y_nR_2}_{y_nR_1} \\
	&+\frac{1}{ \log(\frac{R_1}{R_2} )y_n^2} \left[r Y_1(r) \log\left(\frac{r}{y_n}\right) + Y_0(r)\right]^{y_nR_2}_{y_nR_1} \nonumber \\
	&+ \frac{Y_0(y_n R_1)\log(R_2)}{\mathcal{J}_0(y_n R_1)y_n^2 \log(\frac{R_1}{R_2})} \left[ r \mathcal{J}_1(r) \right]^{y_n R_2}_{y_n R_1} - \frac{\log(R_2)}{y_n^2 \log(\frac{R_1}{R_2})} \left[ r Y_1(r) \right]^{y_n R_2}_{y_n R_1}.\nonumber
	\end{align}
	Note that there may be no (computational) advantage by using these explicit solutions above since these integrals can be evaluated numerically rather fast. Hence it may be convenient to use build in functions of the preferred CAS to find solutions. We recommend this especially for the Equations \eqref{t_2} and \eqref{t_3}.
	
\end{document}